\documentclass[reprint,amsmath,amssymb,aps,prb,floatfix,nolongbibliography,superscriptaddress]{revtex4-2}
\usepackage{iftex}
\ifPDFTeX
  \usepackage[T1]{fontenc}
  \usepackage[utf8]{inputenc}
  \usepackage{lmodern}
\else
  \usepackage{fontspec}
\fi
\usepackage{graphicx}
\usepackage{bm}
\usepackage[unicode,hidelinks,hypertexnames=false]{hyperref}
\newcommand{\Tr}{\operatorname{Tr}}

\newcommand{\ee}{\mathrm{e}}
\newcommand{\iu}{\mathrm{i}}
\newcommand{\unitlambda}{\mathrm{eV}\,\text{\AA}^{-2}\,\mathrm{amu}^{-1}}
\begin{document}

\title{A stochastic propagation method for large-scale phonon calculations}
\author{Wenyingdi He}
\affiliation{Key Laboratory of Artificial Micro- and Nano-Structures of Ministry of Education and School of Physics and Technology, Wuhan University, Wuhan 430072, China}
\author{Weiqing Zhou}
\affiliation{Key Laboratory of Artificial Micro- and Nano-Structures of Ministry of Education and School of Physics and Technology, Wuhan University, Wuhan 430072, China}
\affiliation{Wuhan Institute of Quantum Technology, Wuhan 430206, China}
\author{Shengjun Yuan}
\email{s.yuan@whu.edu.cn}
\affiliation{Key Laboratory of Artificial Micro- and Nano-Structures of Ministry of Education and School of Physics and Technology, Wuhan University, Wuhan 430072, China}
\affiliation{Wuhan Institute of Quantum Technology, Wuhan 430206, China}
\affiliation{School of Artificial Intelligence, Wuhan University, Wuhan 430072, China}
\date{\today}
\begin{abstract}
Large-scale phonon calculations for structurally complex materials are limited by the cubic cost of full diagonalization. We extend the stochastic propagation method (sPM) from electronic systems to lattice dynamics for estimating phonon frequencies, constructing vibrational densities of states, and reconstructing selected phonon polarization vectors by propagating random initial vectors under the dynamical matrix. For finite-range interactions, the leading propagation cost scales linearly with atom number at fixed sampling and propagation parameters. Our error analysis shows that, for nondegenerate modes that are spectrally resolved and have nonzero frequency in the small-error regime, extending propagation yields faster asymptotic reduction of frequency-estimation errors than increasing the number of random initial states alone. Comparisons with direct diagonalization validate the spectral estimates and a selected polarization vector. Convergence tests support the predicted scaling of frequency errors for selected resolved modes.
\end{abstract}
\maketitle

\section{Introduction}

Phonons are fundamental excitations of solids which provide the microscopic basis to understand thermodynamical properties, heat transport and electron-lattice interactions~\cite{Baroni2001,Giustino2017}. Understanding how disorder and structural heterogeneity modify collective atomic vibrations often requires atomistic models large enough to account for defect distributions, interfaces or long-period superstructures~\cite{Hu2021Multiscale,Liu2022Moire}. Thus, large-scale phonon calculations allow a quantitative assessment of these structural features in terms of vibrational spectra, dynamical stability and vibrational contributions to thermodynamic properties~\cite{Togo2023Review}. Relating the vibrational frequencies to mode shapes is a prerequisite for understanding heat transport in structurally complex materials.

Harmonic phonon calculations depend on interatomic force constants, which describe the interactions that govern small vibrations of atoms. The resulting dynamical matrix gives access to phonon frequencies and polarization vectors. Machine-learning interatomic potentials are more efficient to evaluate forces than first-principles methods~\cite{Behler2015,He2022}. Packages like hiphive~\cite{Eriksson2019} and symfc~\cite{Seko2024} incorporate regression and symmetry based methods to enhance the efficiency of force constant determination. These advances permit constructing dynamical matrices for larger systems, but the extraction of corresponding phonon properties can still be computationally expensive. Full diagonalization is therefore increasingly restrictive for the large atomistic models needed to describe disorder and structural complexity~\cite{Kriuchevskyi2020}.

Alternative computational strategies have been developed for large-scale electronic-structure and lattice dynamical calculations. Representative linear-scaling methods in electronic-structure theory exploit the locality of the density matrix or use expansions of the Fermi operator~\cite{Goedecker1999,Zhou2026Review}. Divide-and-conquer methods divide the system into subsystems~\cite{Yang1991}. Stochastic density-functional theory predicts electronic observables using stochastic orbitals~\cite{Baer2013}. One approach in lattice dynamics is to couple localized electronic state calculations and a real space cutoff of the dynamical matrix with maximum entropy reconstruction of the vibrational spectrum~\cite{Ordejon1995}. The kernel polynomial method (KPM) employs Chebyshev expansions and stochastic trace estimation to determine vibrational densities of states and nonaffine viscoelastic response in amorphous solids~\cite{Weisse2006,Kriuchevskyi2020}. For thermal transport, a Kubo approach in real space combines stochastic trace estimation and Chebyshev propagation under the dynamical matrix with Lanczos recursion to compute frequency-dependent phonon diffusion coefficients and thermal conductance~\cite{Li2011}. Reconstruction of eigenvectors by frequency filtering of propagated states was established for Schr\"odinger eigenproblems by Feit et al.~\cite{Feit1982}.

Random-state propagation has been developed in great detail for electronic structure, especially for time domain stochastic propagation methods~\cite{Hams2000,Yuan2010,Zhou2023,Zhou2026Review}. In the tight-binding propagation method (TBPM), time correlation functions, through Fourier transforms, provide electronic densities of states, and superpositions of propagated states pick up states at a given energy and build quasi-eigenstates that keep spatial information at that energy~\cite{Yuan2010}. The propagation time can be increased in a systematic way to improve the energy resolution~\cite{Zhou2026Review}. The density functional propagation method (DFPM) uses time integration to suppress oscillatory cross terms in stochastic estimates of the electron density, reducing sampling noise for local observables~\cite{Zhou2023,Zhou2026Review}. These capabilities motivate the adaptation of time-domain stochastic propagation for large scale phonon calculations.

Here we extend sPM from electronic systems to phonon calculations enabling vibrational densities of states, phonon frequencies, and selected polarization vectors to be obtained without full diagonalization. In contrast to electronic propagation under a Hamiltonian, the random states here evolve under the dynamical matrix in mass weighted coordinates. Frequencies are estimated from crossings of the integrated density of states at half integer values, and polarization vectors are reconstructed by superposing the propagated states with weights that depend on frequency. For finite-range couplings, the propagation cost scales linearly with the number of atoms at fixed sampling and propagation parameters. We analyze errors in individual phonon frequency estimates by separating the finite-duration shift in the absence of sampling fluctuations from the additional shift introduced by finite random-state sampling. For spectrally resolved, nondegenerate modes at nonzero frequency in the small-error regime, the frequency shift in the absence of sampling fluctuations has an approximately $T^{-2}$ envelope with Hamming windowing. Finite random-state sampling introduces an additional shift whose root-mean-square magnitude scales as $R^{-1/2}T^{-1}$, with propagation duration $T$ and random-state number $R$. Since the leading propagation cost grows linearly with either parameter, these local scalings favor extending propagation over increasing sampling alone. Comparisons with direct diagonalization validate phonon dispersions and a selected mode polarization vector in SrTiO$_3$. Tests on silica support the predicted time dependences for selected resolved modes and assess numerical accuracy, computational cost, and large-system vibrational densities of states.

The paper is structured as follows. Section~\ref{sec:method} describes the propagation method, the reconstruction of phonon polarization vectors, the main frequency-error estimates and the computational cost. Section~\ref{sec:results} presents numerical validation and computational benchmarks. The findings and conclusion of the paper are given in Section~\ref{sec:conclusion}. The detailed error analyses are shown in Appendices~\ref{app:sampling} and \ref{app:window} and the numerical propagation is described in Appendix~\ref{app:implementation}.

\section{Stochastic propagation method for lattice dynamics}\label{sec:method}

We formulate a stochastic propagation method for obtaining vibrational densities of states, phonon frequency estimates, and selected phonon polarization vectors without full diagonalization. In the numerical tests, we denote this method by sPM and use direct diagonalization (Diag) as the reference.

\subsection{Phonon frequencies and vibrational density of states}

\begingroup
\predisplaypenalty=0
\brokenpenalty=100
In the harmonic approximation, interatomic force constants describe the couplings between atomic displacements. For a periodic structure, the resulting mass-weighted dynamical matrix is
\begin{equation}
 D_{\kappa\alpha,\kappa'\beta}(\mathbf q)=
 \frac{1}{\sqrt{m_\kappa m_{\kappa'}}}
 \sum_l\Phi_{0\kappa\alpha,l\kappa'\beta}
 \ee^{\iu\mathbf q\cdot(\mathbf R^0_{l\kappa'}-\mathbf R^0_{0\kappa})}.
 \label{eq:dynamical}
\end{equation}
Here, the harmonic force constants are denoted by $\Phi$, the atomic mass is denoted by $m_\kappa$, $\alpha,\beta$ are Cartesian components and $\mathbf R^0_{l\kappa}$ is the equilibrium position of atom $\kappa$ in cell $l$. We include the basis-atom positions in the phase convention. At a fixed wave vector, $D$ is an $M\times M$ Hermitian matrix, with $M=3N$ for $N$ atoms in the chosen cell~\cite{Togo2023Review}.
\par\endgroup

The eigenvalues and orthonormal eigenvectors satisfy
\begin{equation}
 D\mathbf e_\nu=\lambda_\nu\mathbf e_\nu,\qquad
 \mathbf e_\mu^\dagger\mathbf e_\nu=\delta_{\mu\nu},
 \label{eq:eigenproblem}
\end{equation}
where $\mathbf e_\nu$ is the polarization vector in mass-weighted coordinates and the angular frequency satisfies
\begin{equation}
 \omega_\nu^2=\lambda_\nu.
 \label{eq:signedfreq}
\end{equation}
Positive eigenvalues correspond to stable oscillatory modes, while negative eigenvalues indicate harmonic instabilities of the reference structure. The present formulation propagates and reconstructs the spectrum in $\lambda$, keeping both stable and unstable modes, compared to electronic-state propagation in the energy variable. The estimated eigenvalues are then converted to phonon frequencies. The density of states in \(\lambda\)
\begin{equation}
 \rho_\lambda(\lambda)=\Tr\delta(\lambda I-D)
 =\sum_{\nu=1}^{M}\delta(\lambda-\lambda_\nu)
 \label{eq:dos}
\end{equation}
assigns unit weight to each mode and integrates to $M$. For $\omega>0$, the exact vibrational density of states (vDOS) satisfies $g(\omega)=\sum_{\lambda_\nu>0}\delta(\omega-\sqrt{\lambda_\nu})=2\omega\rho_\lambda(\omega^2)$. These expressions describe the modes at a fixed wave vector, whose index is suppressed below. The total vDOS is obtained by Brillouin-zone averaging.

We generate $R$ independent random initial vectors directly in the mass-weighted atomic displacement basis in which $D$ is represented and normalize them to unit Euclidean norm, without further mass weighting:
\begin{equation}
 \|\mathbf e_a^{(0)}\|_2=1,\qquad
 \mathbb E[\mathbf e_a^{(0)}(\mathbf e_a^{(0)})^\dagger]=I/M.
 \label{eq:random}
\end{equation}
The index $a=1,\ldots,R$ labels independent samples. Equation~\eqref{eq:random} gives an expected initial weight of $1/M$ for each mode, and in a finite sample the weights fluctuate with a variance depending on the distribution of the random vector~\cite{Iitaka2004}.

We propagate each random vector by $\mathbf e_a(\tau)=\ee^{-\iu D\tau}\mathbf e_a^{(0)}$. This auxiliary evolution uses a parameter $\tau$ with units reciprocal to squared frequency, compared to the propagation of harmonic wave packets generated by $\sqrt D$ for stable modes. A link to harmonic wave-packet dynamics can nevertheless be made by the frequency-dependent rescaling $\tau=t/(2\omega)$, $\omega>0$, relating frequency-resolved mean-square displacements within a first-order treatment of the perturbation of $D$~\cite{Li2011}. Here, propagation under $D$ is employed to recover its vibrational spectrum. It is unitary for negative eigenvalues and does not replicate the growth of a physically unstable displacement.

The sampling condition gives $M\mathbb E[(\mathbf e_a^{(0)})^\dagger\mathbf e_a(\tau)]=\Tr\ee^{-\iu D\tau}$. Expanding the trace in phonon modes and Fourier transforming leads to the density of states $\rho_\lambda(\lambda)$:

\begin{equation}
 \begin{aligned}
 \rho_\lambda(\lambda)={}&\frac{M}{2\pi}
 \int_{-\infty}^{\infty}\ee^{\iu\lambda\tau}\\
 &\times\mathbb E\!\left[(\mathbf e_a^{(0)})^\dagger
 \mathbf e_a(\tau)\right]d\tau.
 \end{aligned}
 \label{eq:tracefourier}
\end{equation}
For finite sampling, we use the averaged autocorrelation function $C_R(\tau)=\frac{M}{R}\sum_{a=1}^{R}(\mathbf e_a^{(0)})^\dagger\mathbf e_a(\tau)$, with $C_R(0)=M$ and $\mathbb E[C_R(\tau)]=\Tr\ee^{-\iu D\tau}$. Equation~\eqref{eq:tracefourier} is exact for the delta-function spectrum and does not require the approximation of perturbation discussed above. A finite propagation window broadens the peaks, while finite random sampling makes their weights fluctuate.

Numerically, we subtract a spectral center $c$ and divide by a scale $\Delta_D$, chosen to place the spectrum of $\widetilde D=(D-cI)/\Delta_D$ inside $[-1,1]$. A Chebyshev expansion evaluates the propagated vectors using matrix--vector products. The estimation of the eigenvalue range and the Chebyshev propagation are described in Appendix~\ref{app:implementation}. Using $C_R(-\tau)=C_R(\tau)^*$, samples $\tau_n=n\Delta\tau$ up to $T=N_t\Delta\tau$ give
\begin{equation}
 \widehat\rho_{R,T}(\lambda)=\frac{\Delta\tau}{2\pi}
 \left[C_R(0)+2\Re\sum_{n=1}^{N_t}w_n C_R(\tau_n)
 \ee^{\iu\lambda\tau_n}\right].
 \label{eq:discrete}
\end{equation}
For the sampled correlation function we use the Hamming window $w_n=0.54+0.46\cos(\pi n/N_t)$~\cite{Harris1978}. The spectral resolution depends on the propagation time.

Instead of looking for individual spectral maxima we estimate phonon frequencies by looking for half-integer crossings of the integrated density of states (IDOS) in $\lambda$. the IDOS of the reconstructed spectrum is

\begin{equation}
 \widehat I_{R,T}(\lambda)=\int_{\lambda_L}^{\lambda}
 \widehat\rho_{R,T}(u)\,du.
 \label{eq:idos}
\end{equation}
Here $\lambda_L$ is the lower boundary of a spectral period containing the eigenvalue range. The numerical integral over this period is normalized to the total mode count $M$. In the exact-spectrum limit, each mode contributes one unit to the IDOS. 

Specifically, the condition
\begin{equation}
 \widehat I_{R,T}(\lambda)=j-\tfrac12,
 \qquad j=1,\ldots,M,
 \label{eq:crossing}
\end{equation}
is used to select eigenvalue estimates $\widehat\lambda_j$, which are converted to frequencies using Eq.~\eqref{eq:signedfreq}. A finer grid in $\lambda$ can improve the numerical determination of these crossings without increasing the spectral resolution. At finite sampling and propagation duration, however, accurate determination of the eigenvalue depends on the fluctuation in the IDOS, spectral resolution, and crossing-selection rule.

The vDOS is then constructed directly from the estimated positive frequencies using a common normalized frequency-broadening function:
\begin{equation}
 \widehat g_\sigma(\omega)=
 \sum_{\widehat\lambda_j>0} B_\sigma(\omega-\widehat\omega_j),
 \qquad \int_{-\infty}^{\infty}B_\sigma(x)\,dx=1.
 \label{eq:vdos_estimate}
\end{equation}
Here we use the normalized Gaussian $B_\sigma(x)=(\sqrt{2\pi}\sigma)^{-1}\exp[-x^2/(2\sigma^2)]$, where $\sigma$ is the standard deviation in the same units as $\omega$. This construction reuses the frequency estimates and assigns equal weight to each included mode.

\subsection{Reconstruction of phonon polarization vectors}

We propagate a random initial vector and filter the propagated vectors to reconstruct the phonon polarization vector at the target frequency. We use the time-domain reconstruction strategy developed in electronic propagation methods~\cite{Feit1982,Yuan2010,Fu2026Concurrent}. In continuous notation, the reconstructed vector at the target eigenvalue $\lambda_*$ and the reconstruction kernel $K_{T_f}(x)$ are defined by
\begin{align}
 \mathbf z_{\lambda_*}&=\frac{1}{2\pi}\int_{-T_f}^{T_f}
 w_f(\tau)\ee^{\iu\lambda_*\tau}\mathbf e_a(\tau)\,d\tau,
 \label{eq:filter}\\
 K_{T_f}(x)&=\frac{1}{2\pi}\int_{-T_f}^{T_f}
 w_f(\tau)\ee^{\iu x\tau}\,d\tau.
 \label{eq:filterkernel}
\end{align}
Here $T_f$ sets the reconstruction interval and $w_f(\tau)$ is the window function. Expansion in phonon modes gives
\begin{equation}
 \mathbf z_{\lambda_*}=\sum_\nu
 (\mathbf e_\nu^\dagger\mathbf e_a^{(0)})
 K_{T_f}(\lambda_*-\lambda_\nu)\mathbf e_\nu.
 \label{eq:filteredexpansion}
\end{equation}
The kernel $K_{T_f}$ weights modes near the target, with characteristic width of order $T_f^{-1}$ in $\lambda$; finite-width tails give contributions from other modes.

When $\mathbf z_{\lambda_*}\ne0$, the normalized reconstructed vector is $\widetilde{\mathbf e}_{\lambda_*}=\mathbf z_{\lambda_*}/\|\mathbf z_{\lambda_*}\|_2$. If the target is sufficiently close to an isolated resolved eigenvalue and that mode dominates the reconstruction, the normalized vector approximates its polarization vector up to an overall phase, preserving relative component amplitudes and phases. If the modes are degenerate or unresolved, the result contains a linear combination of their polarization vectors instead. A complete basis in a degenerate subspace is not typically determined by a single reconstruction.

Given an initial state, polarization vectors at different target eigenvalues can be reconstructed from the same propagated vectors by changing the target-dependent phase factors in Eq.~\eqref{eq:filter}. The related computational and storage costs are discussed in Sec.~\ref{sec:cost}.

\subsection{Accuracy of phonon frequency estimates}\label{sec:analysis}

The frequency estimates are affected by finite propagation duration, random sampling, and numerical implementation. Let $I_{R,T}$ denote the IDOS obtained from $R$ random initial states at propagation duration $T$, evaluated with exact propagation and integration at the chosen numerical parameters. The associated density is $\rho_{R,T}=dI_{R,T}/d\lambda$. The ensemble-averaged IDOS, $I_T^0=\mathbb E[I_{R,T}]$, assigns unit weight to every mode. The corresponding eigenvalue estimates are defined by the selected solutions of $I_{R,T}(\lambda_j^c)=j-\tfrac12$ and $I_T^0(\lambda_j^0)=j-\tfrac12$, respectively. Relative to the exact eigenvalue $\lambda_j$, the numerical error separates as
\begin{equation}
 \widehat\lambda_j-\lambda_j=b_j(T)+\varepsilon_j(R,T)+\eta_j,
 \label{eq:decomposition}
\end{equation}
where $b_j=\lambda_j^0-\lambda_j$ is the sampling-free finite-duration shift evaluated from the ensemble-averaged spectrum, $\varepsilon_j=\lambda_j^c-\lambda_j^0$ is the sampling-induced shift introduced by a finite number of random initial states at the same propagation duration, and $\eta_j=\widehat\lambda_j-\lambda_j^c$ contains the remaining propagation and numerical-extraction errors. Both $b_j(T)$ and $\varepsilon_j(R,T)$ depend on propagation duration; the latter also depends on random-state sampling. These contributions need not be statistically independent.

Finite random sampling causes fluctuations in the mode weights and hence in the IDOS, shifting the estimated eigenvalues. Expanding to first order in the sampling-induced changes in mode weights gives
\begin{equation}
 \varepsilon_j\simeq-\frac{\delta I_{R,T}(\lambda_j^0)}
 {\rho_T^0(\lambda_j^0)},\qquad
 \delta I_{R,T}=I_{R,T}-I_T^0,\quad \rho_T^0=\frac{dI_T^0}{d\lambda}.
 \label{eq:inverse}
\end{equation}
This estimate requires sufficiently small weight fluctuations and an eigenvalue shift small compared with the width of the same resolved mode peak. For independent initial states, root-mean-square (RMS) stochastic trace-estimation errors decrease as $R^{-1/2}$~\cite{Iitaka2004}. Applying the linearized crossing condition gives the same dependence for the RMS sampling-induced eigenvalue shift at fixed dynamical matrix, propagation duration, and window, as derived in Appendix~\ref{app:sampling}.

The conversion from eigenvalues to frequencies makes low-frequency modes particularly sensitive. For a positive eigenvalue and $|\delta\lambda_j|\ll\lambda_j$,
\begin{equation}
 \delta\omega_j=\frac{\delta\lambda_j}{2\omega_j}
 +O\!\left(\frac{\delta\lambda_j^2}{\omega_j^3}\right).
 \label{eq:frequencyerror}
\end{equation}
For a negative eigenvalue, the imaginary-frequency magnitude $\gamma_j=\sqrt{-\lambda_j}$ instead changes as $\delta\gamma_j\simeq-\delta\lambda_j/(2\gamma_j)$. For near-zero eigenvalues or estimates that change sign, the full square-root conversion must be retained.

To isolate the effect of finite propagation duration, we first consider the ensemble-averaged spectrum, in which random mode-weight fluctuations have been removed. Longer propagation narrows the broadened mode contributions and reduces their overlap. We consider the regime in which further reducing the propagation step $\Delta\tau$ at fixed $T$ no longer appreciably changes the estimated eigenvalues. For a resolved, nondegenerate mode with nearest eigenvalue separation $d_j$ and $Td_j\gg1$, the Hamming-window analysis gives
\begin{equation}
 |b_j(T)|_{\rm env}=O\!\left(\frac{1}{T^2d_j}\right),
 \label{eq:envelope}
\end{equation}
when the sampling-free finite-duration shift is dominated by contributions from a few modes closest in eigenvalue, under the conditions detailed in Appendix~\ref{app:window}. Here the subscript ``env'' denotes the envelope of the oscillatory error magnitude $|b_j(T)|$.

For a given set of random initial states, increasing propagation does not change the mode weights, but narrows the peak corresponding to each resolved mode. This means that a fluctuation in the IDOS produces a smaller shift in the estimated eigenvalue. Within the first-order treatment of the sampling-induced weight changes discussed above, the RMS eigenvalue error for estimates based on $R$ independent random initial states satisfies
\begin{equation}
 \operatorname{RMS}(\varepsilon_j)\sim C_j R^{-1/2}T^{-1},
 \label{eq:randomTscale}
\end{equation}
The coefficient $C_j$ depends on the mode spectrum and sampling statistics. Their derivations are given in Appendix~\ref{app:window}.

For a given dynamical matrix, propagation step and propagation accuracy the leading propagation cost scales linearly with $R$ or $T$. In the resolved-mode regime discussed above, increasing the propagation produces a $T^{-2}$ envelope for the sampling-free finite-duration shift and a $T^{-1}$ RMS sampling-induced shift for fixed $R$, both decaying faster than the $R^{-1/2}$ reduction obtained from increasing the number of random states alone. For a fixed mode of nonzero frequency with small eigenvalue error Eq.~\eqref{eq:frequencyerror} maps these time dependences onto the frequency error. These estimates are for individual resolved modes and do not give a universal convergence exponent for the full spectrum root-mean-square error (RMSE).

\begin{figure*}[!tp]
 \centering
 \begin{minipage}[t]{.48\textwidth}\vspace{0pt}\centering
 \includegraphics[width=\linewidth]{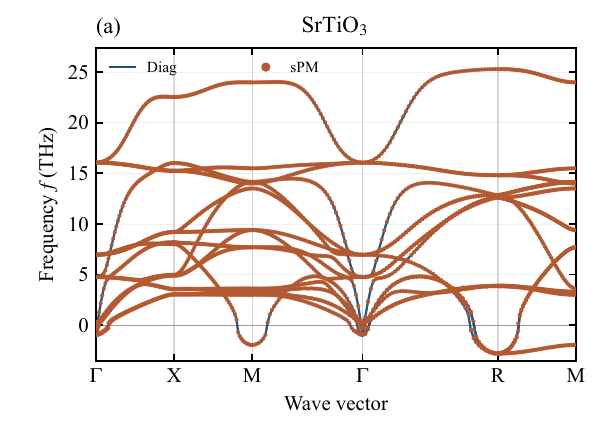}
 \end{minipage}\hfill
 \begin{minipage}[t]{.50\textwidth}\vspace{0pt}\centering
 \makebox[.48\linewidth][l]{\small (b)}\hfill\makebox[.48\linewidth][l]{\small (c)}\par
 \makebox[.48\linewidth][c]{\small\sffamily Diagonalization}\hfill\makebox[.48\linewidth][c]{\small\sffamily sPM polarization vector}\par
 \includegraphics[width=\linewidth,trim=0 0 0 19.2bp,clip]{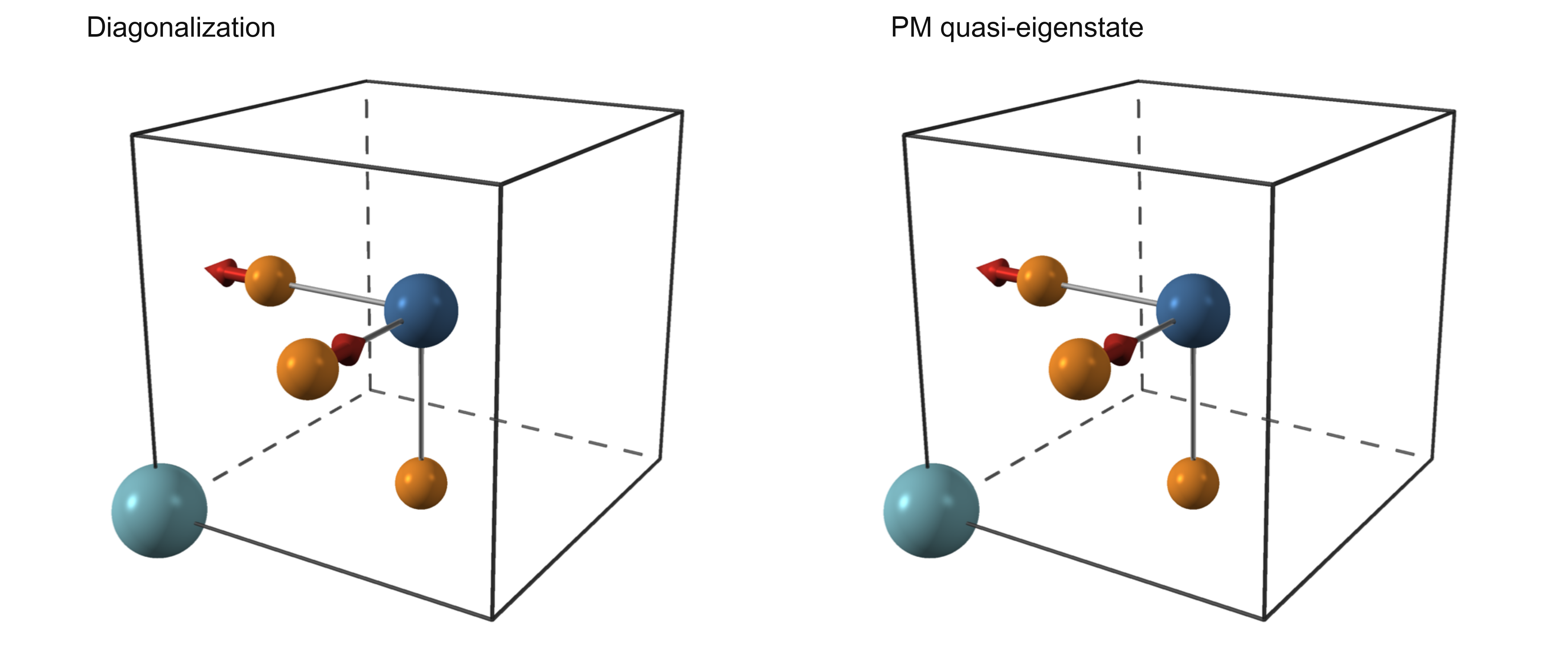}
 \end{minipage}
 \caption{Phonon dispersion and a selected phonon polarization vector of cubic SrTiO$_3$. (a) Frequencies obtained by direct diagonalization (Diag, blue lines) and sPM (brown circles) from the same dynamical matrices, using $R=160$ and $N_t=1024$. Imaginary frequencies are displayed as negative values. (b) Diag polarization vector and (c) normalized sPM polarization vector for mode 14 of a separate M-point validation matrix. Arrows show phase-aligned real vector components in one reference primitive cell, with the same scale in both panels. Cyan, blue, and orange spheres represent Sr, Ti, and O, respectively.}
 \label{fig:sto}
\end{figure*}

\subsection{Computational cost for large systems}\label{sec:cost}

For force constants of finite effective range, each atomic displacement couples to a spatially limited set of atoms. Let $N_{\mathrm{nz}}$ denote the total number of nonzero elements in $D$, and $z$ their average number per row, so that $N_{\mathrm{nz}}=zM$. At fixed density and effective coupling range, $z$ is approximately independent of system size~\cite{Kriuchevskyi2020}. This locality also occurs in short-range machine-learning interatomic potentials~\cite{Behler2015}. Explicitly retained long-range interactions require separate treatment.

As in DFPM, matrix--vector products are the dominant operations in the time propagation~\cite{Zhou2023}. We denote the computational cost of propagating $R$ random initial vectors for $N_t$ steps by $C_{\mathrm{prop}}$. Its leading operation count scales as
\begin{equation}
 C_{\rm prop}=O(RN_tzN),
 \label{eq:cost}
\end{equation}
at fixed rescaled time step and Chebyshev accuracy, so that the number of matrix--vector products per step remains fixed. The cost therefore grows linearly with atom number for fixed $R$, $N_t$, and $z$. This estimate concerns propagation after matrix construction. Force evaluation, force-constant construction, spectrum reconstruction, and frequency estimation incur additional costs. Resolving individual eigenvalues in increasingly dense spectra can require longer propagation. However, thermal conductivity can be evaluated from correlation functions without resolving every individual mode~\cite{Li2011,Fan2019HNEMD}. The required propagation duration should therefore be determined by convergence of the target observable.

For a sequential evaluation of random states, propagation itself requires only a small number of vectors of length $M$, in addition to the stored matrix, scalar correlations, and spectral arrays. Once the target eigenvalues have been estimated, multiple polarization vectors can be reconstructed from the same stored propagation trajectory using target-dependent phase factors. For a fixed number of stored trajectory samples and target polarization vectors, the trajectory and reconstruction-vector storage grows linearly with system size.

Stochastic trace-estimation theory provides a reference for the size dependence of spectral sampling errors~\cite{Hams2000,Iitaka2004}. For normalized Gaussian or random-phase initial states at fixed spectral broadening, the root-mean-square error of the trace-based spectrum normalized by the number of modes satisfies
\begin{equation}
 \sigma_{\rm stat}=O\!\left((RM)^{-1/2}\right)
 =O\!\left((RN)^{-1/2}\right).
 \label{eq:size_selfaveraging}
\end{equation}
Because the normalized spectral error scales as $(RN)^{-1/2}$, larger systems require fewer random states for a fixed normalized spectral accuracy.

\section{Numerical results}\label{sec:results}

First we test phonon dispersions and a particular polarization vector in crystalline SrTiO$_3$, then frequency convergence, the large-system vDOS, and computational scaling in amorphous silica. An in-house code includes displacement generation, the force-evaluation workflow, the construction of the force-constant and dynamical-matrix, and sPM calculations. Numerical reference for each comparison is obtained by direct diagonalization of the same matrices.

\begin{figure*}[!tp]
 \centering
 \includegraphics[width=.88\textwidth]{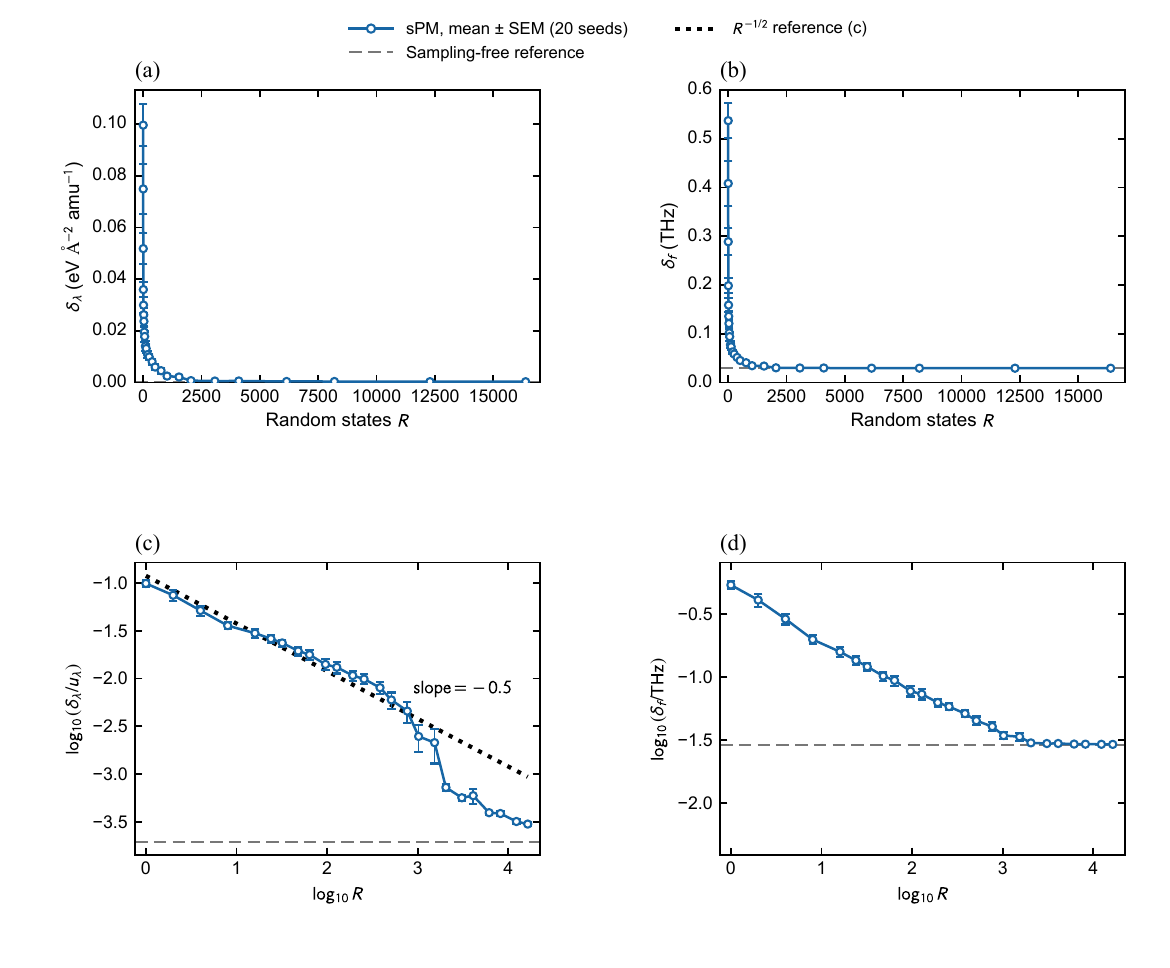}
 \caption{Dependence of eigenvalue and phonon-frequency errors on the number of random states for the 243-atom silica model. (a,b) All-mode eigenvalue and frequency RMSEs relative to Diag; (c,d) the corresponding logarithmic plots. Calculations use $N_t=2048$. Each point is the mean RMSE over 20 independent sets of $R$ random initial states; error bars indicate the standard error of the mean. The dotted line in (c) indicates $R^{-1/2}$ scaling. Dashed horizontal lines show the sampling-free reference errors obtained from the exact Diag trace using the same finite window and numerical extraction procedure. Here $u_\lambda=\unitlambda$ is the eigenvalue unit used in (c).}
 \label{fig:sampling}
\end{figure*}

\begin{figure*}[!tp]
 \centering
 \begin{minipage}{.485\textwidth}\centering
 \includegraphics[width=\linewidth]{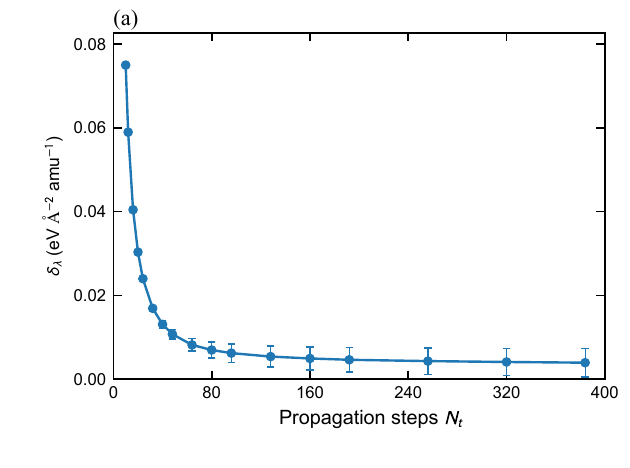}\end{minipage}\hfill
 \begin{minipage}{.485\textwidth}\centering
 \includegraphics[width=\linewidth]{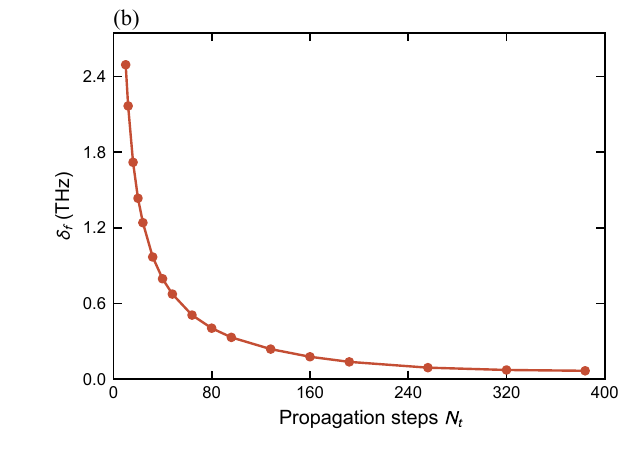}\end{minipage}
 \par\smallskip
 \begin{minipage}{.485\textwidth}\centering
 \includegraphics[width=\linewidth]{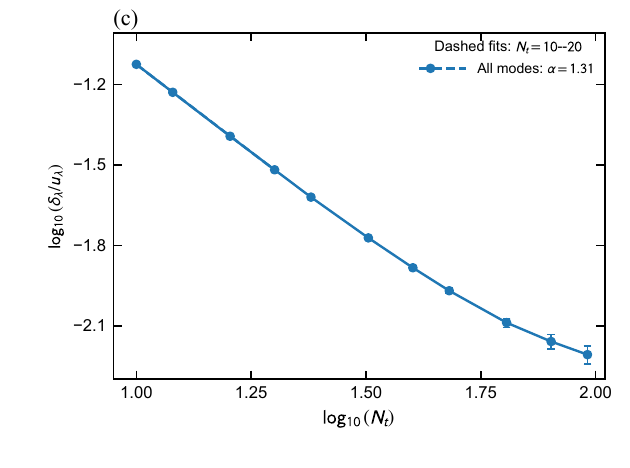}\end{minipage}\hfill
 \begin{minipage}{.485\textwidth}\centering
 \includegraphics[width=\linewidth]{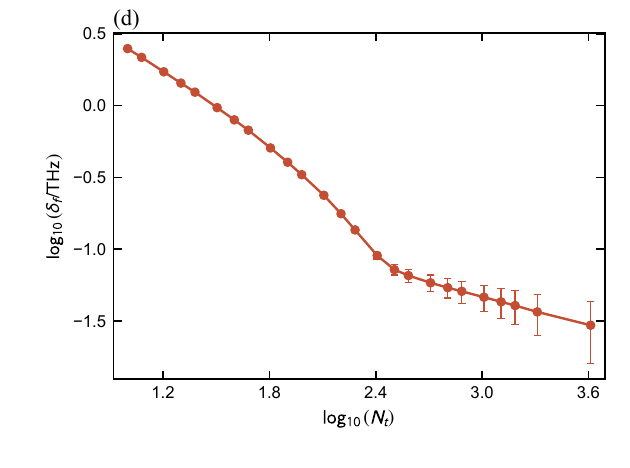}\end{minipage}
 \caption{Eigenvalue and phonon-frequency errors versus the number of propagation steps for the 243-atom silica model, with $R=1024$. (a,b) All-mode eigenvalue and frequency RMSEs relative to Diag; (c,d) Eigenvalue and frequency errors shown on logarithmic axes. Points are mean RMSEs over 20 independent random-seed groups. Error bars show the standard deviation across the 20 sets in (a,b,d). In (c), they show the standard error of the mean. The dashed line in (c) is a fit to $N_t=10,12,16,20$, giving $\delta_\lambda\propto N_t^{-1.31}$.}
 \label{fig:duration}
\end{figure*}

\begin{figure*}[!tp]
 \centering
 \includegraphics[width=.84\textwidth]{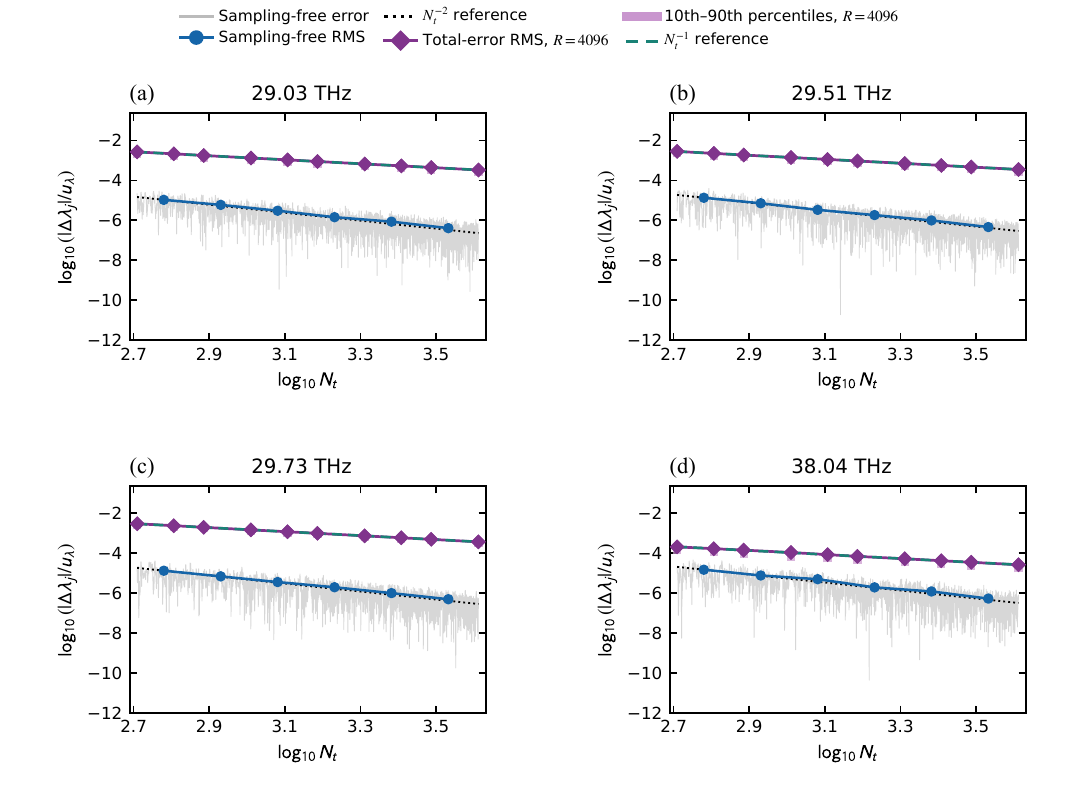}
 \caption{Propagation-length dependence of eigenvalue errors for four resolved modes of the 243-atom silica model. The panel titles give the Diag frequencies. Grey curves show sampling-free finite-duration errors obtained from the exact Diag trace. Blue circles show the RMS of the sampling-free errors over successive intervals of $N_t$. Purple diamonds show the RMS total error over five groups of $R=4096$ random initial states. Pale-purple bars indicate the 10th--90th percentile range of absolute group errors. Eigenvalues are estimated from continuous half-integer crossings of the IDOS after Hamming windowing. The black dotted and green dashed lines indicate fixed-exponent $N_t^{-2}$ and $N_t^{-1}$ references, respectively. Errors are shown on a logarithmic scale with $u_\lambda=\unitlambda$.}
 \label{fig:finite_time_errors}
\end{figure*}

\begin{figure*}[!tp]
 \centering
 \begin{minipage}[t]{.485\textwidth}\centering
 \includegraphics[width=\linewidth]{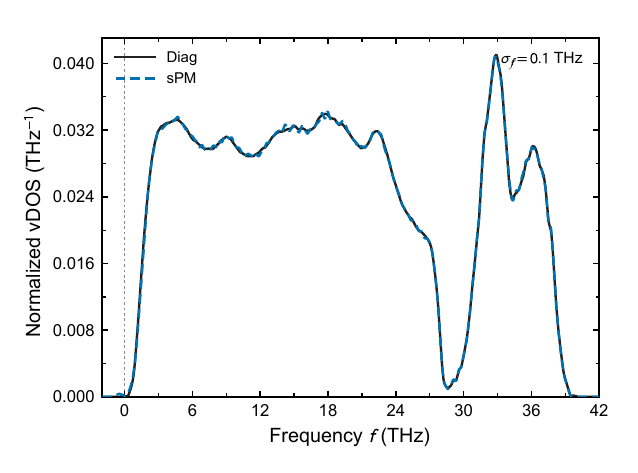}
 \caption{Vibrational density of states of the $36\,864$-atom amorphous silica model at the supercell $\Gamma$ point. Black solid and blue dashed curves are obtained from Diag and sPM frequencies, respectively. Both curves are normalized to unit area and broadened with a Gaussian of standard deviation $\sigma_f=0.1$ THz. sPM uses $R=15$ and $N_t=1024$.}
 \label{fig:large_silica_dos}
 \end{minipage}\hfill
 \begin{minipage}[t]{.485\textwidth}\centering
 \includegraphics[width=\linewidth]{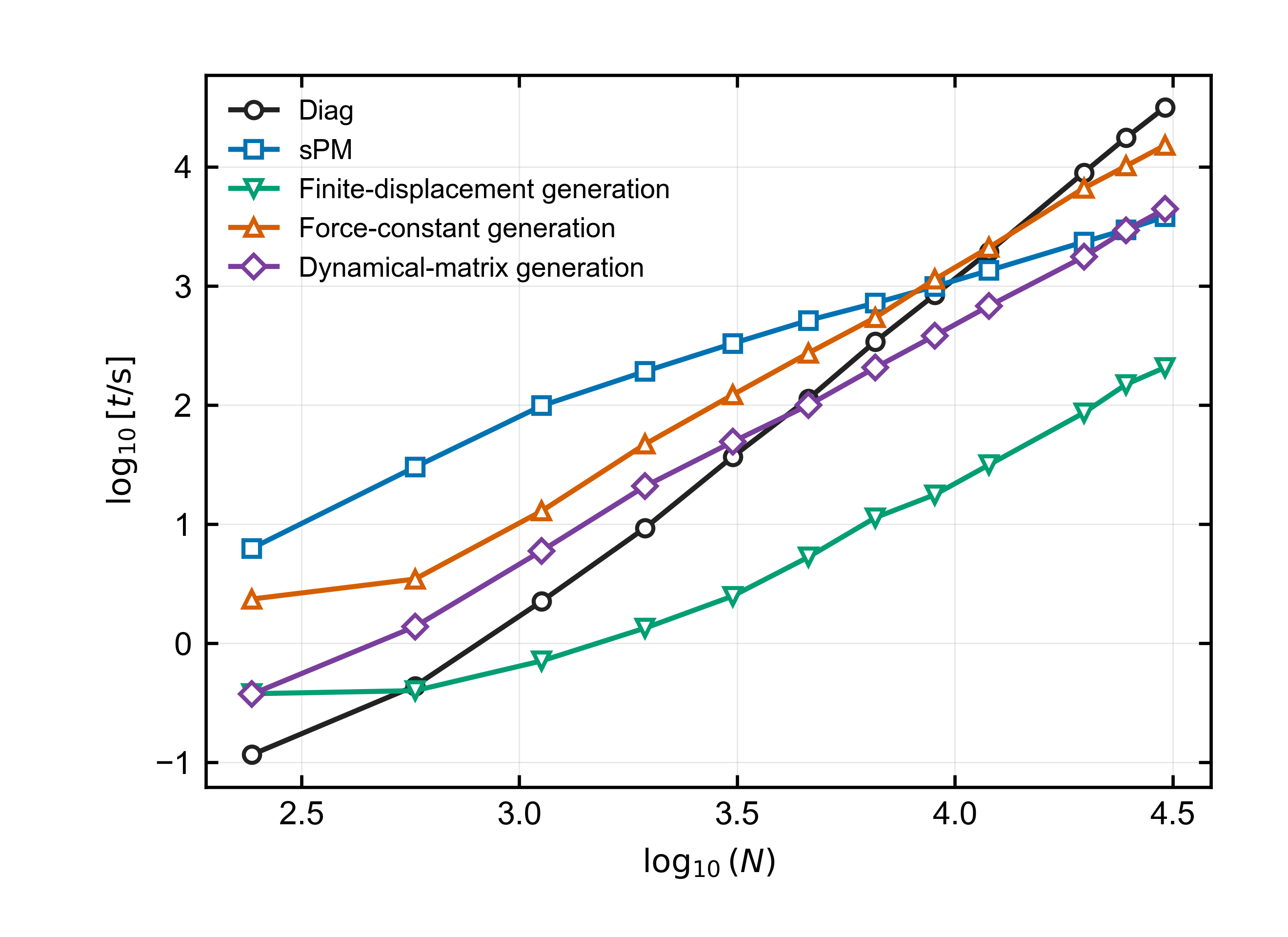}
 \caption{Wall-clock times for individual stages of phonon calculations in 12 amorphous silica models containing 243--$30\,375$ atoms. The stages are identified in the legend; dynamical-matrix timings include matrix output. sPM uses $R=15$ and $N_t=1024$ throughout. Both axes show base-10 logarithms, with time measured in seconds. Lines are guides to the eye.}
 \label{fig:timing}
 \end{minipage}
\end{figure*}

\subsection{Phonon dispersion of cubic \texorpdfstring{SrTiO$_3$}{SrTiO3}}\label{sec:stodispersion}

SrTiO$_3$ (STO) is a prototypical perovskite oxide with competing polar and antiferrodistortive lattice instabilities. In bulk STO, the cubic-to-tetragonal transition near 105 K involves antiphase rotations of neighboring TiO$_6$ octahedra; the low-temperature tetragonal phase is an incipient ferroelectric~\cite{He2022}. These distinct structural tendencies make cubic STO a useful benchmark for recovering both stable and unstable harmonic phonon branches.

We use the Deep Potential model trained on PBEsol data~\cite{He2022}, with a relaxed cubic lattice constant of $3.896$ \AA. Force constants are obtained from a $4\times4\times4$ supercell containing 320 atoms. No additional acoustic-sum-rule or nonanalytic corrections are applied. The five-atom primitive cell gives 15 branches at each wave vector, so this calculation tests spectral accuracy rather than large-matrix performance.

Figure~\ref{fig:sto} (a) shows a comparison of sPM and Diag at 251 wave vectors along $\Gamma$--X--M--$\Gamma$--R--M. Hamming windowing is used and the spacing of the uniform eigenvalue grid is $\Delta\lambda_{\mathrm{grid}}=\pi/(128T)$. The frequencies are given as $f=\omega/(2\pi)$ in THz. The imaginary frequencies are represented as negative for the plots and frequency-error statistics. sPM reproduces the dispersion over the entire path. The all-branch frequency RMSE is $0.0274$ THz against Diag. Largest discrepancies are at near zero frequency, where the conversion from eigenvalues to frequencies is especially sensitive as discussed in Sec.~\ref{sec:analysis}.

sPM reproduces the unstable branches near $\Gamma$ and R which are related to the polar and octahedral-rotation tendencies respectively. The recovery of these branches tests the ability to extract instability related spectral information without discarding negative eigenvalues. These are harmonic properties of the cubic reference structure and not finite-temperature, anharmonically renormalized phonon frequencies~\cite{Tadano2015}.

\subsection{Reconstruction of a selected phonon polarization vector}\label{sec:stopolarization}

We next test whether propagation recovers the atomic components of a selected mode, in addition to its frequency. Figures~\ref{fig:sto}(b,c) use an isolated stable mode of a separate STO M-point validation matrix at $\mathbf q=(0.5,0.5,0)$ in reciprocal-lattice units. Phonon frequencies are estimated using $R=20$ random initial states propagated for 512 steps. The polarization vector is then reconstructed in a separate propagation of 256 steps. The Diag frequency is $17.9916$ THz, compared with a sPM estimate of $17.9893$ THz. For normalized Diag and sPM vectors, we evaluate the squared overlap and eigenvalue residual,
\begin{align}
 P_j&=|\mathbf e_j^\dagger\widetilde{\mathbf e}_j|^2,\nonumber\\
 r_j&=\|(D-\widehat\lambda_j I)\widetilde{\mathbf e}_j\|_2.
 \label{eq:modechecks}
\end{align}
The reconstructed mode gives $P_{14}=0.99987$ and $r_{14}=1.13\times10^{-2}\,\unitlambda$, using the sPM eigenvalue estimate in the residual.

The vibration is dominated by two oxygen atoms with principal components along the $x$ and $y$ directions, respectively, and much smaller Sr and Ti components. sPM reproduces their relative amplitudes and directions, showing that the polarization vector retains the atomic vibration pattern rather than only a spectral peak. A future extension could use independent random-state propagations and singular-value decomposition of the reconstructed vectors to obtain a basis for a degenerate phonon subspace.

\subsection{Convergence of phonon frequency estimates}\label{sec:samplingresults}\label{sec:timeresults}\label{sec:resolvedresults}

The convergence tests use a fixed $\Gamma$-point dynamical matrix of a melt-quenched silica configuration containing 81 Si and 162 O atoms. Its $M=729$ modes allow direct comparison with the complete Diag spectrum. For each independent random-state group, we calculate
\begin{align}
 \delta_\lambda&=\left[\frac1M\sum_{j=1}^M(\widehat\lambda_j-\lambda_j)^2\right]^{1/2},\nonumber\\
 \delta_f&=\left[\frac1M\sum_{j=1}^M(\widehat f_j-f_j)^2\right]^{1/2},
 \label{eq:rmse}
\end{align}
pairing modes in eigenvalue order and including all modes in the errors. Figures~\ref{fig:sampling} and \ref{fig:duration} report mean RMSEs over 20 independent groups generated with different random seeds, allowing us to assess the robustness of the convergence trends against random-state sampling for this silica configuration.

Increasing $R$ at fixed propagation duration reduces the fluctuations in mode weights. In Fig.~\ref{fig:sampling}, the mean frequency RMSE decreases from $0.537$ THz at $R=1$ to $0.0293$ THz at $R=16384$. The $R^{-1/2}$ reference illustrates the sampling dependence discussed in Sec.~\ref{sec:analysis}, but the total frequency error approaches a plateau. The reference obtained from the exact Diag trace, processed with the same finite window and readout, is $0.0292$ THz. The remaining frequency error at large $R$ is dominated by the near-zero modes, consistent with their enhanced sensitivity to eigenvalue errors. Increasing the random-state count alone therefore cannot remove errors due to finite propagation duration in the absence of sampling fluctuations or errors due to numerical extraction.

Longer propagation narrows the mode contributions and improves the frequency estimates at fixed $R$. Figure~\ref{fig:duration} shows this behavior for $R=1024$: the mean frequency RMSE falls from $2.49$ THz at $N_t=10$ to $0.507$ THz at $N_t=64$ and $0.0296$ THz at $N_t=4096$. The four-point fit in panel (c) gives an effective eigenvalue-error exponent of $1.31$ over the selected early interval. As discussed in Sec.~\ref{sec:analysis}, the estimated eigenvalues are affected by finite propagation duration, random mode-weight fluctuations, and numerical extraction. The all-mode RMSE combines these errors over both resolved and unresolved parts of the spectrum. Evaluating the same finite-window reconstruction with the exact Diag trace gives nearly the same early decrease over $N_t=10$--20, indicating that sampling-free finite-duration effects dominate this interval. The fitted exponent therefore characterizes finite-range convergence of the full spectrum rather than the asymptotic behavior of an isolated mode.

To examine these distinct contributions, Fig.~\ref{fig:finite_time_errors} considers four resolved modes at longer propagation durations. We first replace the random-state correlation by the exact trace evaluated from the Diag eigenvalues, thereby removing sampling fluctuations. Continuous half-integer crossings of the IDOS isolate the sampling-free finite-duration shift $b_j(T)$. The RMS of these shifts over successive step-count intervals follows approximately the $T^{-2}$ reference, consistent with Eq.~\eqref{eq:envelope}. Restoring finite random-state sampling gives larger total errors whose RMS across groups decreases approximately along the $T^{-1}$ reference, as expected when the sampling-induced shift dominates, consistent with Eq.~\eqref{eq:randomTscale}. At fixed $\Delta\tau$, these time dependences are equivalent to dependences on $N_t$.

Here $\Delta\lambda_j=\lambda_j^{\rm est}-\lambda_j^{\rm Diag}$, and $512\leq N_t\leq4096$ with the same $\Delta\tau$ as in the preceding scans. The $N_t^{-2}$ reference is anchored to the first blue RMS point; only the amplitude of the $N_t^{-1}$ reference is fitted.

These resolved-mode tests distinguish sampling-free finite-duration shifts from sampling-induced shifts in the estimated eigenvalues, but do not quantitatively determine the early all-mode convergence exponent. When neighboring modes are unresolved, their broadened contributions overlap and modify the IDOS used to estimate their frequencies. Longer propagation narrows these contributions and improves frequency discrimination, whereas increasing $R$ at fixed duration reduces weight fluctuations without improving spectral resolution. For resolved, nondegenerate modes at nonzero frequency in the small-error regime of Sec.~\ref{sec:analysis}, the sampling-free finite-duration shift has a $T^{-2}$ envelope, while the RMS sampling-induced shift decreases as $T^{-1}$ at fixed $R$. Both dependences are faster than the $R^{-1/2}$ reduction obtained by increasing the number of random states alone. Since the leading propagation cost grows linearly with either $T$ or $R$, extending propagation provides an efficient route to improved frequency accuracy beyond increasing the number of random states alone.

\subsection{Large-scale vibrational spectra and computational scaling}\label{sec:largedos}\label{sec:timing}

We next apply the frequency-reconstruction procedure to a dense amorphous silica model containing $36\,864$ atoms and $110\,592$ modes. The model is prepared with the BKS potential~\cite{vanBeest1990}, supplemented by a short-range 24--6 correction~\cite{Guissani1996} and a finite-cutoff Wolf treatment of electrostatics~\cite{Wolf1999}. A periodic melt is quenched from 3000 to 300 K over 1 ns and subsequently relaxed. The resulting density is approximately $2.57$ g/cm$^3$, with a network dominated by fourfold-coordinated Si and twofold-coordinated O. The supercell $\Gamma$-point calculation samples the vibrations of this periodic amorphous network without assuming a smaller crystalline primitive cell. This system size is chosen to permit a direct comparison with full diagonalization. It represents a validation scale rather than an intrinsic size limit of the propagation method, whose fixed-parameter cost scales approximately linearly with system size.

Fig.~\ref{fig:large_silica_dos} shows vDOS curves calculated from the Diag and sPM frequencies with a common Gaussian width of $0.1$ THz. The sPM calculation uses the same eigenvalue-grid spacing $\Delta\lambda_{\mathrm{grid}}=\pi/(128T)$. The spectrum exhibits broad lower frequency bands, a low weight region around 28--30 THz and a higher frequency doublet around 33 and 36 THz. Previous mode-resolved studies of amorphous silica associate the high-frequency region with Si--O stretching vibrations, whereas the lower-frequency modes show stronger bending and collective-network character~\cite{Bhattarai2016}. sPM reproduces the broad bands and the doublet obtained by direct diagonalization, maintaining the main spectral features of this amorphous silica model.

The area between the sPM and Diag curves is $0.00624$ after normalizing each vDOS to unit area, indicating good agreement in the broadened spectral distribution. This comparison validates the reconstruction of a smooth vDOS from the estimated frequencies, complementing the individual-frequency tests above. The Gaussian width specifies numerical smoothing rather than a physical vibrational linewidth. The total vDOS characterizes the distribution of harmonic vibrations; selected polarization vectors provide access to their atomic vibration patterns.

Having established the accuracy of the large-system vDOS, we next examine how the computational cost grows with system size. The silica size series uses independently melt-quenched periodic models and the same finite-cutoff interaction model described above. The finite interaction range limits the number of atoms coupled to each displacement, providing the setting for the near-linear propagation cost in Sec.~\ref{sec:cost}. Figure~\ref{fig:timing} compares 12 sizes from 243 to $30\,375$ atoms with $R=15$ and $N_t=1024$ fixed. The benchmarks were configured with 32 OpenMP threads.

The sPM runtime grows substantially more slowly with system size than the Diag runtime, consistent with the fixed-parameter cost estimate in Eq.~\eqref{eq:cost}. At $30\,375$ atoms, the Diag and sPM stages take $3.16\times10^4$ and $3.81\times10^3$ s, respectively. The sPM stage is therefore approximately 8.3 times faster than direct diagonalization, illustrating the computational advantage of propagation for the tested large-system spectral calculation.

Force-constant and dynamical-matrix construction remain important parts of the calculation. At the largest plotted size, force-constant generation takes approximately $1.52\times10^4$ s, exceeding the sPM stage. The advantage of propagation therefore concerns obtaining phonon information from an available dynamical matrix; further reductions in total workflow cost also depend on efficient force evaluation and matrix construction.

\section{Discussion and conclusions}\label{sec:conclusion}

We have extended sPM from electronic systems to lattice dynamics in large systems that estimates phonon frequencies, constructs vibrational densities of states, and reconstructs selected phonon polarization vectors without full diagonalization. Frequencies are estimated from the IDOS and selected polarization vectors are reconstructed from superpositions of propagated states. For an isolated, resolved mode, the reconstruction recovers its relative atomic amplitudes and phases. The framework thus connects vibrational frequencies with the atomic patterns of selected modes.

Comparisons with direct diagonalization assess these complementary outputs. The STO dispersion calculation recovers stable and unstable phonon branches, while a separate M-point validation matrix tests the polarization vector of an isolated mode. Silica calculations examine frequency convergence with random-state number and propagation duration and recover the broad bands and high-frequency doublet of a large-system vDOS. These tests establish the method's accuracy for the spectral and mode information examined here.

The error analysis provides a practical basis for improving frequency accuracy. Under the resolved-mode conditions of Sec.~\ref{sec:analysis}, the eigenvalue shift in the absence of sampling fluctuations has a $T^{-2}$ envelope with Hamming windowing. Finite random-state sampling introduces an additional shift whose RMS scales as $R^{-1/2}T^{-1}$. For nonzero-frequency modes with sufficiently small errors, these eigenvalue-error scalings also apply to frequency estimates. Longer propagation leaves the random mode weights unchanged but narrows their spectral peaks, reducing the frequency shifts caused by fluctuations in the IDOS. The selected-mode silica tests support these time dependences. At fixed propagation step and accuracy, the leading propagation cost grows linearly with either $R$ or $T$; in this regime, extending propagation therefore offers a more favorable error reduction than increasing the random-state count alone.

For finite-range couplings at fixed density, sampling and propagation parameters the leading propagation cost scales linearly with the number of atoms. The silica timing series shows a slower growth in runtime for sPM than for full diagonalization at the sizes tested. This cost estimate is for propagation after dynamical-matrix construction. In practice, the resolution needed depends on the observable targeted: resolving each individual frequency or reconstructing a specific polarization vector does not require the same precision as convergence of a broadened vDOS. The selection of the propagation duration is important for the applicability of the method to large systems with dense vibrational spectra.

Extending sPM to lattice dynamics provides a spectral and mode basis for investigating microscopic heat transport in structurally complex solids. Selected polarization vectors characterize the spatial extent of vibrations and provide projections of atomic motion onto individual resolved modes. Combining these projections with molecular dynamics trajectories and heat current calculations provides a route to determining how different vibrations contribute to energy transport~\cite{Larkin2014}. This connection is central to understanding how disorder redistributes heat among vibrational modes and how vibrations near defects or interfaces affect energy transfer. These questions link the physics of disordered solids to the control of lattice thermal conductivity in thermoelectric materials and heat dissipation in nanoscale devices. The present method establishes the spectral foundation for pursuing these problems in large atomistic models.

\begin{acknowledgments}
This work was supported by the National Natural Science Foundation of China (Grants No. 12425407 and 12547165). W.Z. gratefully acknowledges support from the Hubei Provincial Young Science and Technology Talent ``Morning Light'' Support Program and the Hongyi postdoctoral fellowship of Wuhan University. We thank the Core Facility of Wuhan University for providing the computational resources.
\end{acknowledgments}

\appendix

\section{Random-state sampling errors in phonon frequency estimates}\label{app:sampling}

We keep the dynamical matrix, propagation duration, time step, and window fixed and consider the error caused by a finite number of random initial states. The reference integrated density of states (IDOS), $I_T^0=\mathbb E[I_{R,T}]$, retains the same finite-duration broadening but averages over the initial states. The reference estimate $\lambda_j^0$ and the random-state estimate $\lambda_j^c$ satisfy
\begin{equation}
 I_T^0(\lambda_j^0)=I_{R,T}(\lambda_j^c)=j-\tfrac12.
 \label{eq:samplingcrossings}
\end{equation}
Their difference is the sampling-induced shift $\varepsilon_j=\lambda_j^c-\lambda_j^0$ in Eq.~\eqref{eq:decomposition}.

With $\delta I_{R,T}=I_{R,T}-I_T^0$ and $\rho_T^0=dI_T^0/d\lambda$, expansion to first order in the random weight fluctuations gives
\begin{equation}
 \begin{aligned}
 0&\simeq\delta I_{R,T}(\lambda_j^0)
       +\rho_T^0(\lambda_j^0)\varepsilon_j,\\
 \varepsilon_j&\simeq-
 \frac{\delta I_{R,T}(\lambda_j^0)}{\rho_T^0(\lambda_j^0)}.
 \end{aligned}
 \label{eq:crossingexpansion}
\end{equation}
This approximation requires a nonzero reference density and a shift small enough that the density changes little between the two estimates. The product of the sampling-induced density fluctuation and the eigenvalue shift is second order in the random weight fluctuations and is neglected. The denominator is therefore evaluated using the reference density.

Let $I_T^{(a)}$ be the IDOS from one normalized initial state, including the factor $M$ that restores the total mode number. Since the IDOS depends linearly on the autocorrelation,
\begin{equation}
 I_{R,T}(\lambda)=\frac1R\sum_{a=1}^{R}I_T^{(a)}(\lambda),
 \qquad \mathbb E[I_T^{(a)}(\lambda)]=I_T^0(\lambda).
 \label{eq:single_state_average}
\end{equation}
At $\lambda_j^0$, define $\delta I_j=\delta I_{R,T}(\lambda_j^0)$ and the single-state fluctuation $\xi_j^{(a)}=I_T^{(a)}(\lambda_j^0)-I_T^0(\lambda_j^0)$. Each $\xi_j^{(a)}$ has zero mean and variance $\sigma_{I,j}^2=\mathbb E[(\xi_j^{(a)})^2]$. For independent initial states drawn from the same distribution, the terms with $a\ne b$ vanish on averaging, leaving $R$ equal contributions:
\begin{equation}
 \begin{aligned}
 \mathbb E[\delta I_j^2]
 &=\frac1{R^2}\sum_{a,b=1}^{R}
   \mathbb E[\xi_j^{(a)}\xi_j^{(b)}]\\
 &=\frac{\sigma_{I,j}^2}{R}.
 \end{aligned}
 \label{eq:countvariance}
\end{equation}
Combining Eqs.~\eqref{eq:crossingexpansion} and \eqref{eq:countvariance} gives
\begin{equation}
 \sqrt{\mathbb E[\varepsilon_j^2]}
 \simeq\frac{\sigma_{I,j}}{|\rho_T^0(\lambda_j^0)|\sqrt R}.
 \label{eq:variance}
\end{equation}
At fixed matrix and propagation parameters, the coefficient is independent of $R$. The root of the mean-square sampling error over a fixed set of modes satisfying these conditions has the same $R^{-1/2}$ dependence.

For $\omega_j^0=\sqrt{\lambda_j^0}>0$ and $|\varepsilon_j|\ll\lambda_j^0$, the frequency difference is $\sqrt{\lambda_j^c}-\sqrt{\lambda_j^0}\simeq\varepsilon_j/(2\omega_j^0)$, so it inherits the same sampling dependence. Sampling-free finite-duration shifts and numerical-extraction errors remain separate contributions to the total error.

\section{Propagation-time dependence of eigenvalue errors}\label{app:window}

\subsection{Sampling-free finite-duration shifts of the estimated eigenvalues}

Finite propagation broadens the contribution of each mode to the density of states in $\lambda$. We determine the resulting sampling-free finite-duration shift $b_j=\lambda_j^0-\lambda_j$ from the IDOS condition in Eq.~\eqref{eq:samplingcrossings}, with the dynamical matrix fixed and $T=N_t\Delta\tau$. At the exact eigenvalue, the IDOS deviation without sampling fluctuations is
\begin{equation}
 A_j(T)=I_T^0(\lambda_j)-(j-\tfrac12).
 \label{eq:duration_IDOS_offsets}
\end{equation}
Since $I_T^0(\lambda_j+b_j)=j-\tfrac12$, the shift satisfies
\begin{equation}
 0=A_j+\int_{\lambda_j}^{\lambda_j+b_j}\rho_T^0(u)\,du.
 \label{eq:countbalance}
\end{equation}
When $b_j$ is small compared with the width of the target mode peak and the density changes little over this displacement, Eq.~\eqref{eq:countbalance} gives
\begin{equation}
 b_j\simeq-\frac{A_j}{\rho_T^0(\lambda_j)}.
 \label{eq:duration_local_shifts}
\end{equation}
The reference density must be nonzero at $\lambda_j$.

\subsection{Finite-duration error without sampling fluctuations}

We consider a mode separated from its neighbors at the chosen propagation duration. After convergence with respect to $\Delta\tau$, we approximate the time sum by an integral over $[-T,T]$ and neglect contributions from the boundaries and periodic copies of the reconstruction interval. For spectral reconstruction, we use the kernel in Eq.~\eqref{eq:filterkernel} with $T_f=T$ and $w_f(\tau)=w(\tau)$. Its contribution to the IDOS is
\begin{equation}
 F_T(y)=\int_{-\infty}^{y}K_T(u)\,du,
 \label{eq:continuouskernel}
\end{equation}
where $w(\tau)=0.54+0.46\cos(\pi\tau/T)$ and $u$ is an eigenvalue difference. Symmetry and normalization give $F_T(0)=1/2$ and $F_T(+\infty)=1$.

Direct integration yields
\begin{equation}
 K_T(u)=\frac{\sin(Tu)}{\pi}
 \left[\frac{0.54}{u}
 -\frac{0.46u}{u^2-(\pi/T)^2}\right].
 \label{eq:hammingkernel}
\end{equation}
The central peak height is $K_T(0)=c_HT$, where $c_H=0.54/\pi$. For $T|u|\gg1$, expansion of the denominator gives, with $0.08=w(\pm T)$,
\begin{equation}
 K_T(u)=\frac{0.08\sin(Tu)}{\pi u}
 +O\!\left(\frac{1}{T^2|u|^3}\right).
 \label{eq:hamming_density_tail}
\end{equation}
Let $y=\lambda-\lambda_\nu$, and let $\Theta(y)$ denote the unit step. For $y>0$, $1-F_T(y)=\int_y^\infty K_T(u)\,du$. Substituting Eq.~\eqref{eq:hamming_density_tail} requires the following integral. With $v=Tu$, integration by parts, followed by a second integration by parts to estimate the remainder, gives
\begin{equation}
 \begin{aligned}
 \int_y^\infty\frac{\sin(Tu)}{u}\,du
 &=\frac{\cos(Ty)}{Ty}
 -\int_{Ty}^\infty\frac{\cos v}{v^2}\,dv\\
 &=\frac{\cos(Ty)}{Ty}+O\!\left((Ty)^{-2}\right).
 \end{aligned}
 \label{eq:hamming_tail_integral}
\end{equation}
Using $F_T(-y)=1-F_T(y)$, we obtain, for $T|y|\gg1$,
\begin{equation}
 F_T(y)-\Theta(y)
 =-\frac{0.08\cos(Ty)}{\pi Ty}
 +O\!\left((T|y|)^{-2}\right).
 \label{eq:hammingcumulativetail}
\end{equation}
For ordered, distinct eigenvalues, define $\Delta_{j\nu}=\lambda_j-\lambda_\nu$. Using $F_T(0)=1/2$ in Eq.~\eqref{eq:duration_IDOS_offsets} gives
\begin{equation}
 \begin{aligned}
 A_j(T)\simeq{}&\sum_{\nu<j}[F_T(\Delta_{j\nu})-1]
 +\sum_{\nu>j}F_T(\Delta_{j\nu})\\
 \simeq{}&-\frac{0.08}{\pi T}
 \sum_{\nu\ne j}\frac{\cos(T\Delta_{j\nu})}{\Delta_{j\nu}}.
 \end{aligned}
 \label{eq:tails}
\end{equation}
When the target mode dominates $\rho_T^0(\lambda_j)$, $\rho_T^0(\lambda_j)=K_T(0)+\sum_{\nu\ne j}K_T(\Delta_{j\nu})\simeq c_HT$. Equation~\eqref{eq:duration_local_shifts} therefore gives
\begin{equation}
 b_j(T)\simeq\frac{0.08}{0.54\,T^2}
 \sum_{\nu\ne j}\frac{\cos(T\Delta_{j\nu})}{\Delta_{j\nu}},
 \label{eq:windowbias}
\end{equation}
and hence
\begin{equation}
 |b_j(T)|_{\rm env}
 =O\!\left(\frac{1}{T^2}
 \sum_{\nu\ne j}\frac{1}{|\Delta_{j\nu}|}\right).
 \label{eq:tailbound}
\end{equation}
Thus $A_j(T)$ has a $T^{-1}$ envelope, while $\rho_T^0(\lambda_j)\propto T$, giving the $T^{-2}$ envelope of $b_j(T)$. If a few nearby modes dominate and $d_j=\min_{\nu\ne j}|\Delta_{j\nu}|$, this reduces to $O(T^{-2}d_j^{-1})$, as in Eq.~\eqref{eq:envelope}. The estimate requires $Td_j\gg1$ and a dominant contribution from the target mode to $\rho_T^0(\lambda_j)$.

\subsection{Propagation-duration dependence of the random-state sampling error}

For a fixed set of $R$ random initial states, the projection onto each phonon mode determines its spectral weight:
\begin{equation}
 p_\nu^{(a)}=|\mathbf e_\nu^\dagger\mathbf e_a^{(0)}|^2,
 \qquad q_\nu=\frac{M}{R}\sum_{a=1}^R p_\nu^{(a)}.
 \label{eq:modeweights}
\end{equation}
The weights obey $\mathbb E[q_\nu]=1$ and $\sum_\nu q_\nu=M$. Propagation changes the phases of the mode components, leaving these weights unchanged. 

In Eq.~\eqref{eq:discrete}, each mode contributes the broadened line shape
\begin{equation}
 K_{T,\Delta\tau}(u)=\frac{\Delta\tau}{2\pi}
 \left[1+2\sum_{n=1}^{N_t}w_n\cos(un\Delta\tau)\right],
 \label{eq:periodickernel}
\end{equation}
where $T=N_t\Delta\tau$ and $w_n=0.54+0.46\cos(\pi n/N_t)$. Its integral over one period $L=2\pi/\Delta\tau$ is one. In the selected interval $[\lambda_L,\lambda_L+L]$, a mode contributes
\begin{equation}
 F_{T,\Delta\tau}(\lambda;\lambda_\nu)
 =\int_{\lambda_L}^{\lambda}K_{T,\Delta\tau}(u-\lambda_\nu)\,du
 \label{eq:periodiccumulative}
\end{equation}
to the IDOS. Thus $I_{R,T}(\lambda)=\sum_\nu q_\nu F_{T,\Delta\tau}(\lambda;\lambda_\nu)$, and setting all $q_\nu=1$ gives $I_T^0$.

The deviation from the expected IDOS at $\lambda_j^0$ is
\begin{equation}
 \delta I_j(R,T)=\sum_\nu(q_\nu-1)
 F_{T,\Delta\tau}(\lambda_j^0;\lambda_\nu).
 \label{eq:random_IDOS_kernel}
\end{equation}

Under the resolved-mode approximation of the preceding subsection, the single-mode IDOS contributions approach one below the target and zero above it. Since $b_j=O(T^{-2})$, the target contribution at $\lambda_j^0$ approaches $1/2$. Thus
\begin{equation}
 \delta I_j(R,T)\simeq
 \sum_{\nu<j}(q_\nu-1)+\tfrac12(q_j-1)
 \equiv Z_j.
 \label{eq:fixedweightcount}
\end{equation}
This remaining IDOS fluctuation is set by the initial states and does not decrease by extending their propagation.

Longer propagation increases the density near the resolved mode in proportion to $T$. When that mode dominates the sampled density, $\rho_{R,T}(\lambda_j^0)\simeq q_jc_HT$. Expanding the IDOS about $\lambda_j^0$ gives $0\simeq Z_j+q_jc_HT\varepsilon_j$, and hence
\begin{equation}
 \varepsilon_j(R,T)\simeq-\frac{Z_j}{q_jc_HT}.
 \label{eq:randomtimebridge}
\end{equation}
The sampling-induced shift therefore decreases as $T^{-1}$ for fixed initial states. This approximation requires a nonzero target weight and a shift small compared with the finite-duration width of the same mode, $T|\varepsilon_j|\simeq|Z_j|/(q_jc_H)\ll1$.

Combining the sampling result of Eq.~\eqref{eq:variance} in Appendix~\ref{app:sampling} with $\rho_T^0(\lambda_j^0)\simeq c_HT$ gives, to first order in the random weight fluctuations,
\begin{equation}
 \operatorname{RMS}(\varepsilon_j)
 \simeq C_j R^{-1/2}T^{-1}.
 \label{eq:randomtimevariance}
\end{equation}
Here $C_j$ is independent of $R$ and $T$ in this resolved-mode approximation, for a fixed dynamical matrix and initial-state distribution, as in Eq.~\eqref{eq:randomTscale}. The RMS is taken over independent groups of $R$ initial states. This estimate applies while the estimated eigenvalue remains close to the same resolved, nondegenerate mode.

\section{Numerical propagation}\label{app:implementation}

We calculate the action of the evolution operator by Chebyshev propagation using matrix--vector products as in TBPM and DFPM~\cite{Yuan2010,Zhou2023,Zhou2026Review}. An $8\times8$ real symmetric tridiagonal matrix is obtained by passing a normalized random vector through eight Lanczos steps in the automatic rescaling procedure. Its smallest and largest eigenvalues serve as estimates of the extreme eigenvalues of $D$. Both estimates are pushed outwards by a correction computed from the norm of the final Lanczos residual and the last components of the eigenvectors of the tridiagonal matrix. These are numerical approximations, not guaranteed enclosing bounds. The interval used for propagation must contain the entire spectrum. Define for the obtained lower and upper estimates
\begin{equation}
 \Delta_D=\frac{\lambda_{\rm high}-\lambda_{\rm low}}{2},\qquad
 c=\frac{\lambda_{\rm high}+\lambda_{\rm low}}{2},
 \label{eq:rescaling}
\end{equation}
and $\widetilde D=(D-cI)/\Delta_D$. The propagation step is $\Delta\tau=\pi/\Delta_D$, corresponding to a reconstructed eigenvalue interval of width $2\Delta_D$.

At each step, the state is updated by
\begin{align}
 \mathbf e_a(\tau_{n+1})=\ee^{-\iu c\Delta\tau}\bigg[&J_0(\Delta_D\Delta\tau)\mathbf v_0\nonumber\\
 &+2\sum_{k=1}^{\infty}(-\iu)^kJ_k(\Delta_D\Delta\tau)\mathbf v_k\bigg],
 \label{eq:chebyshev}
\end{align}
where $J_k$ is a Bessel function and $\mathbf v_k=T_k(\widetilde D)\mathbf e_a(\tau_n)$, with $T_k$ the first-kind Chebyshev polynomial. Starting from $\mathbf v_0=\mathbf e_a(\tau_n)$ and $\mathbf v_1=\widetilde D\mathbf v_0$, the recurrence is
\begin{equation}
 \mathbf v_{k+1}=2\widetilde D\mathbf v_k-\mathbf v_{k-1}.
 \label{eq:recurrence}
\end{equation}
Since $\Delta_D\Delta\tau=\pi$ is fixed, the same Bessel coefficients are used at each step. The series is truncated immediately before the first order $k>\pi$ for which $|J_k(\pi)|$ falls below a prescribed coefficient tolerance. Each additional polynomial order requires one matrix--vector product~\cite{Yuan2010}.

\let\originalbibfont\bibfont
\renewcommand{\bibfont}{\originalbibfont\setlength{\baselineskip}{0.96\baselineskip}}
\bibliography{references}
\end{document}